\documentclass[prl,amssymb,twocolumn,showpacs]{revtex4}

\usepackage{graphicx}
\usepackage{times}
\usepackage{psfrag}

\newcommand{\ket}[1]{| #1 \rangle}

\newcommand{\be}{\begin{equation}}
\newcommand{\ee}{\end{equation}}
\newcommand{\bae}{\begin{eqnarray}}
\newcommand{\eae}{\end{eqnarray}}
\newcommand{\ep}{\epsilon}

\def\CC{{\rm\kern.24em \vrule width.04em height1.46ex depth-.07ex
    \kern-.30em C}}
\def\P{{\rm I\kern-.25em P}}

\def\bbbc{{\mathchoice {\setbox0=\hbox{$\displaystyle\rm C$}\hbox{\hbox
to0pt{\kern0.4\wd0\vrule height0.9\ht0\hss}\box0}}
{\setbox0=\hbox{$\textstyle\rm C$}\hbox{\hbox
to0pt{\kern0.4\wd0\vrule height0.9\ht0\hss}\box0}}
{\setbox0=\hbox{$\scriptstyle\rm C$}\hbox{\hbox
to0pt{\kern0.4\wd0\vrule height0.9\ht0\hss}\box0}}
{\setbox0=\hbox{$\scriptscriptstyle\rm C$}\hbox{\hbox
to0pt{\kern0.4\wd0\vrule height0.9\ht0\hss}\box0}}}}

\def\bbbz{{\mathchoice {\hbox{$\sf\textstyle Z\kern-0.4em Z$}}
{\hbox{$\sf\textstyle Z\kern-0.4em Z$}}
{\hbox{$\sf\scriptstyle Z\kern-0.3em Z$}}
{\hbox{$\sf\scriptscriptstyle Z\kern-0.2em Z$}}}}

\begin{document}

\title{Two-Point Versus Multipartite Entanglement in Quantum Phase Transitions}
\author{Alberto Anfossi$^1$, Paolo Giorda$^{1,2}$, Arianna Montorsi$^1$, and Fabio Traversa$^1$}
\affiliation{$^1$ Dipartimento di Fisica del Politecnico, C.so
Duca degli Abruzzi 24, I-10129 Torino, Italy} \affiliation{$^2$
Institute for Scientific Interchange (ISI), Villa Gualino, Viale
Settimio Severo 65, I-10133 Torino, Italy}
\date{July 28, 2005}
\begin{abstract}
We analyze correlations between subsystems for an extended Hubbard
model exactly solvable in one dimension, which exhibits a rich
structure of quantum phase transitions (QPTs). The $T=0$ phase
diagram is exactly reproduced by studying singularities of
single-site entanglement. It is shown how comparison of the latter
quantity and quantum mutual information allows one to recognize
whether two-point or shared quantum correlations are responsible
for each of the occurring QPTs. The method works in principle for
any number $D$ of degrees of freedom per site. As a by-product, we
are providing a benchmark for direct measures of bipartite
entanglement; in particular, here we discuss the role of
negativity at the transition.
\end{abstract}

 \maketitle

In the past years, the characterization of complex quantum
phenomena has received a strong impulse from the recent
developments in quantum-information theory. Within such framework,
a crucial notion is that of entanglement. Besides being recognized
as a fundamental resource for quantum computation and
communication tasks \cite{qip2}, it has also been used to better
characterize the critical behavior of different many-body quantum
systems when some characteristic parameter of the related
Hamiltonian is varied; the latter phenomenon being known as
quantum phase transition (QPT) \cite{sadchso}.
\\In fact, a deep comprehension of universal properties of QPTs
has not  been fully reached yet. The peculiarity of using
entanglement in this context is that, being a single direct
measure of quantum correlations, it should allow for a unified
treatment of QPTs; at least, whenever the occurring QPT is to
ascribe to the quantum nature of the system, which is always the
case at $T=0$ since thermal fluctuations are absent.\\
A first description of the relations between entanglement of one
or two spins and QPTs in spin-$1/2$ chains was given in
\cite{fazioNielsen}, where it was noticed how derivatives of
concurrence show divergencies in correspondence of QPT, with
appropriate scaling exponents. The entanglement of blocks of L
spins and its scaling behavior in spin models showing critical
behavior was then investigated in \cite{LatVidKor}. The problem of
characterizing the ground state phase diagram of fermionic systems
by means of entanglement has been addressed more recently in
\cite{GuliExtHub}, where it was shown how the study of
\textit{single-site entanglement} allows one to reproduce the
relevant features of the known (numerical) phase diagram. While
this is a promising starting point, it remains to be clarified
which quantum correlations are responsible for the occurring QPT:
if two-points or shared (multipartite), if short or long ranged.
The answer to the above issue would in fact require exhaustive
investigation of the entanglement between any two subsystems. In
case the subsystems have just two degrees of freedom, concurrence
properly quantifies the quantum correlations \cite{Woo}. A
generalization of such quantity to (sub)systems with a higher
number of degrees of freedom $D$ has been proposed, and is known
as \textit{negativity} \cite{vidalvernerneg}. Also, the total
amount of correlations between any two subsystems is captured by
\textit{quantum mutual information} \cite{GPW}.
\\In the following we describe a method based on the comparison of
the latter quantities for arguing whether the occurring transition
is to ascribe to two-points or multipartite quantum correlations;
the method works for arbitrary $D$. Our strategy is tested on a
one-dimensional extended Hubbard model that was solved
\cite{AA,SCHAD}, exhibiting a rich structure of phase diagram at
$T=0$. We show that the phase diagram is exactly reproduced by the
singularities of single-site entanglement. We then infer which of
the QPTs is originated from a singular behavior of two-point or
multipartite entanglement; our results are confirmed by the exact
solution.

\paragraph{Correlations and subsystems.} We are interested in the
existing correlations between: $a)$ the single site $i$ and the
rest of the system; $b)$ the generic site $i$ and a generic site
$j\neq i$; $c)$ the generic pair of site $(i,j)$ (dimer) and the
rest of the system.
\\In order to measure the \textit{total correlations} between two generic
subsystems $A$ and $B$ we use the quantum mutual information
\cite{vedralrmp,qip2,GPW}. The latter is defined as
 \be
\mathcal{I}_{AB}=S(\rho_{A})+S(\rho_{B})-S(\rho_{AB}),\label{QMI}
\ee
 where $\rho_{AB}$, $\rho_{A}$ and $\rho_{B}$ are the system's
and subsystems' density matrices, respectively, and $S(\rho)=-\sum
_i \lambda_i \log_2 \lambda_i $ ($\lambda_i$ being the eigenvalues
of $\rho$) is the Von Neumann entropy. In \cite{vedralrmp,GPW} it
was shown how $\mathcal{I}_{AB}$ is a proper measure of all
(quantum and classical) correlations between $A$ and $B$. In case
$A$ and $B$ are single sites, we will refer to the latter as
two-point quantum ($Q2$) and classical ($C2$) correlations.
\\As far as \textit{quantum correlations} are concerned, we
consider two different cases. When $\rho_{AB}$ is a pure state,
correlations between $A$ and $B$ are purely quantum and are
measured by $S(\rho_{A})=S(\rho_{B})$. This happens when $A$
corresponds to a single site $i$  (or to the dimer $i,j$), and $B$
corresponds to the remaining sites \cite{PZFL};
$\mathcal{S}_i=S(\rho_i)$ (single-site entanglement) accounts for
both the localized correlations ($Q2$) and the shared ones ($QS$
in the following). When we deal instead with the correlations
between two generic sites $(i,j)$, the density matrix of the
global system is the dimer's one: $\rho_{AB}=\rho_{ij}$. The
latter generally corresponds to a mixed state. Thus, to evaluate
the quantum correlations between two generic sites, we need a
measure of entanglement for bipartite mixed states. In general,
proposed measures are hard to compute whenever $D>2$, since they
require difficult optimization processes. However, there is at
least one measure easy to compute\cite{vidalvernerneg}, the
negativity
  \be
  \mathcal{N}(\rho_{AB})=(\|\rho_{AB}^{T_A}\|_1-1)/2;\label{neg}
  \ee
 where $\rho_{AB}^{T_A}$ is the partial
 transposition with respect to the subsystem $A$ applied on
 $\rho_{AB}$, and $\|O\|_1\doteq Tr\sqrt{ O^\dagger O}$ is the trace norm
 of the operator $O$. $\rho_{AB}^{T_A}$ can have negative eigenvalues
 $\mu_i$, and the negativity can also be expressed as
$\mathcal{N}(\rho_{AB})=|\sum_i \mu_i|$. Although negativity is
not a perfect measure of entanglement \cite{convexroofneg}, it
gives important bounds for quantum information protocols i.e.,
teleportation capacity and asymptotic distillability. Its role in
describing QPTs has not been fully investigated yet.

\paragraph{Entanglement and QPTs.}
$\mathcal{S}_i$ has been proven to be a useful tool in describing
QPTs \cite{GuliExtHub}. As already pointed out, to give a
better characterization of the latter, one could as well consider
 quantum correlations between different subsystems.\\
The scheme we propose in this letter is based on the idea of
comparing $\mathcal{S}_i$ --not allowing one to distinguish Q2
from QS correlations-- with different functionals quantifying
instead just two-point correlations. We study $\mathcal{N}_{i,j}$,
which is at least a lower bound for $Q2$ correlations, and
$\mathcal{I}_{i,j}$, which properly captures total ($Q2$ and $C2$)
correlations. As a first step, the exact phase diagram is obtained
analyzing the singularities shown by $\mathcal{S}_i$,
$\mathcal{I}_{i,j}$, and $\mathcal{N}_{i,j}$. Successively, a
comparison of the singular behavior of $\mathcal{I}_{i,j}$ with
that of $\mathcal{S}_i$ allows one to discriminate whether a QPT
is to ascribe to $Q2$ or $QS$ correlations. In fact, whenever
$\mathcal{S}_i$ exhibits a singular behavior due to $Q2$
correlations, the \textit{same} type of singular behavior should
be highlighted as well by $\mathcal{I}_{i,j}$ (since it also
contains $Q2$ correlations), and possibly by $\mathcal{N}_{i,j}$,
in case the latter would properly capture them for our model. On
the contrary, when the singular behavior of $\mathcal{S}_i$ is to
ascribe to $QS$ correlations, the same singular behavior should
not be displayed either by $\mathcal{I}_{i,j}$ or by
$\mathcal{N}_{i,j}$, since both measures regard only two-point
correlations.

\paragraph{The bond-charge extended Hubbard model.}
The model we deal with is described by the following Hamiltonian:
\be H_{BC} = u \sum_i n_{i \uparrow}n_{i \downarrow}
-\sum_{<i,j>\sigma} [1 - x (n_{i \bar{\sigma}}+n_{j \bar{\sigma}})
]c_{i \sigma}^\dagger c_{j \sigma} \, , \label {ham_bc} \ee where
$c_{{i} \sigma}^\dagger , c_{{i} \sigma}^{} \,$ are fermionic
creation and annihilation operators on a one-dimensional chain of
length $L$; $\sigma = \uparrow, \downarrow$ is the spin label,
$\bar{\sigma}$ denotes its opposite, ${n}^{}_{j \sigma} = c_{j
\sigma}^\dagger c_{j \sigma}^{}$ is the spin $\sigma$ electron
charge, and $\langle {i} , \, {j} \rangle$ stands for neighboring
sites; $u$ and $x$ ($0\leq x \leq 1$) are the (dimensionless)
on-site Coulomb repulsion and bond-charge interaction parameters.
\begin{figure}[h]
\begin{centering}
\psfrag{u}{\small $u$}
\psfrag{U}{\small $u$}
\psfrag{D}{\small$n$}
 \psfrag{n}{\small $n$}
\psfrag{T}{\small $\partial_u\mathcal{S}_{i}$}
\hbox{\includegraphics[height=4.25cm, width=4.07cm, viewport= 10
10 300 220, clip]{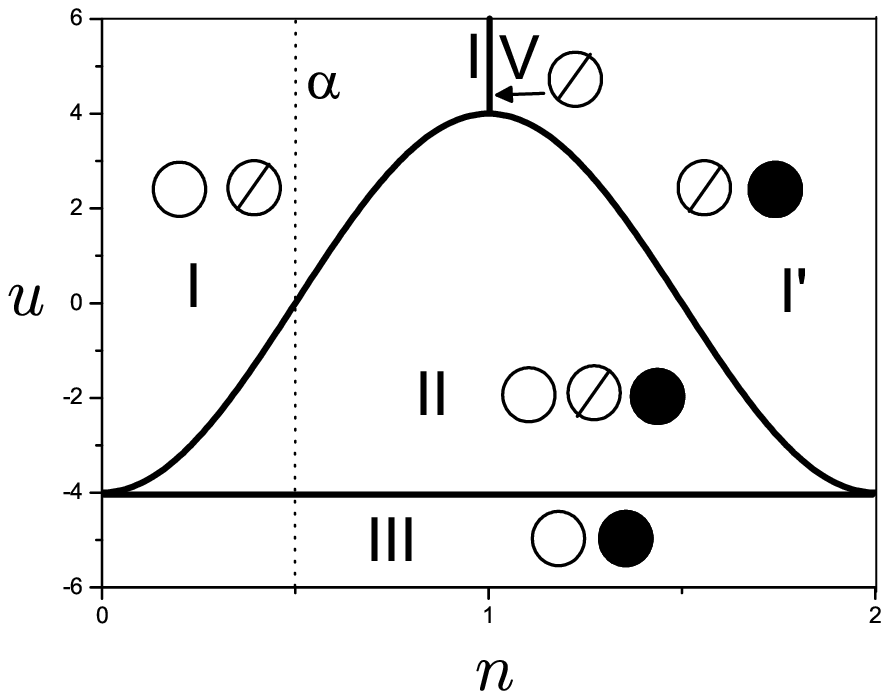} \hfill
\includegraphics[height=4.25cm, width=4.5cm, viewport= 60 250 580 825,
clip]{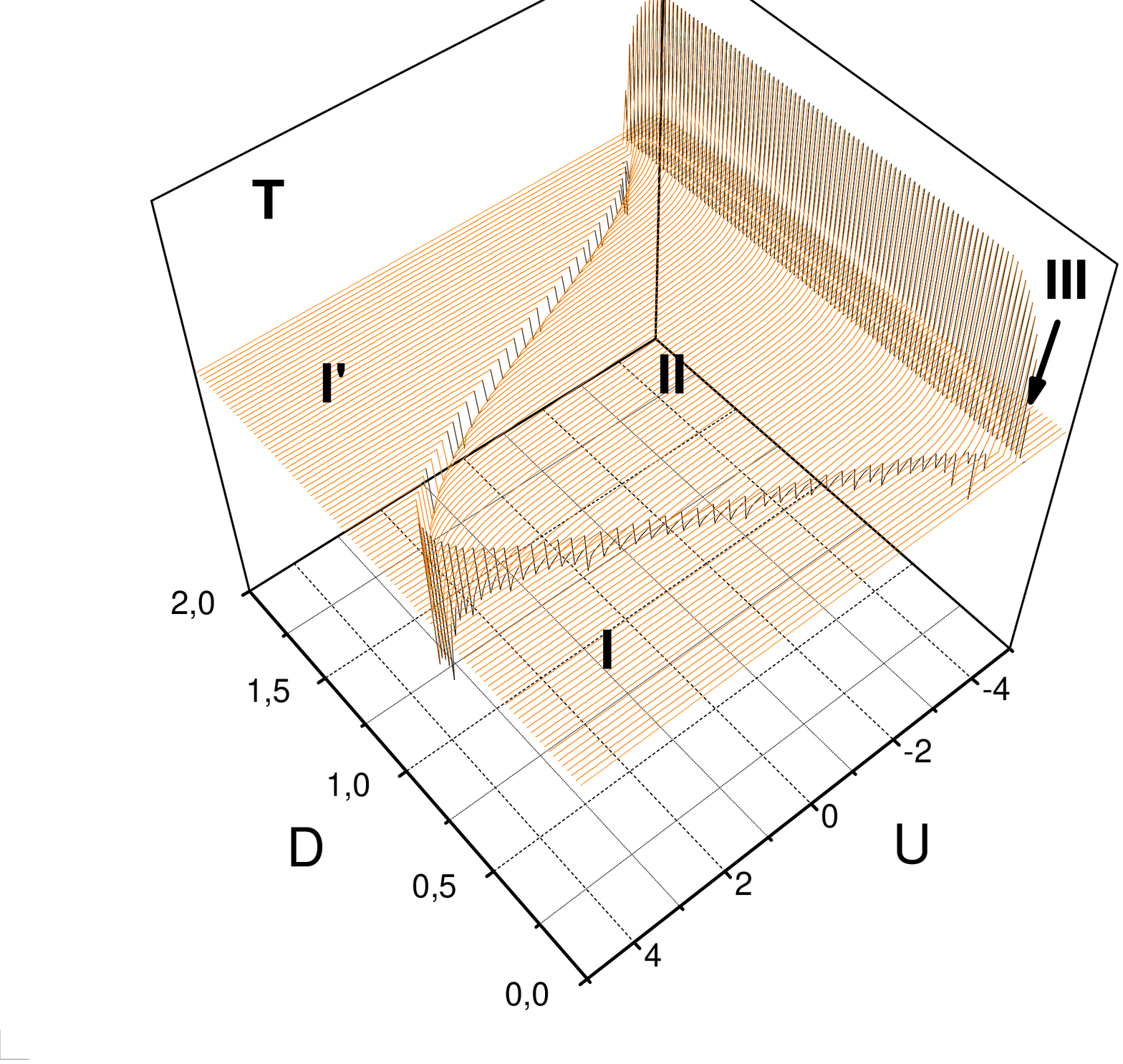}}
 \caption{\small{Left: Ground-state phase diagram.
 Empty, slashed and full dots stay for empty, singly and doubly
 occupied sites. Right: $\partial_u\mathcal{S}_{i}$.}}
 \label{3D}
 \end{centering}
\end{figure}
\\The model is considered here at $x=1$, in which case the number
of doubly occupied sites becomes a conserved quantity. The exact
eigenstates of $H_{BC}(x=1)$ are obtained in \cite{AA,SCHAD,DOMO};
the ground-state phase diagram is shown in the left part of Fig.
\ref{3D}. The latter presents various QPTs driven by parameters
$u$ and average number of electrons per site (filling) $n$. The
charge-gapped phase $IV$ is insulating
 and all sites are singly occupied; phases $I$, $I'$, and $II$ fall in the
Tomonaga-Luttinger class (neither spin nor charge gap); they are
characterized by the presence of singly and empty sites (phase
$I$), singly and doubly occupied sites (phase $I'$), both of which
have dominant charge-charge correlations, and all types of sites
(phase $II$) with superconducting correlations and off-diagonal
long range order (ODLRO). The latter characterizes also phase
$III$, where sites are empty or doubly occupied.\\
The model's energy spectrum is fully independent of spin
orientation \cite{AA}: any sequence of spins in the chain cannot
be altered by the Hamiltonian, which, in fact, acts on a Hilbert
space that at each site $i$ has $D_i=3$, and is spanned by the
states $|0\rangle_i$ (empty), $|\mbox{\o}\rangle_i$ (singly
occupied), and $|2\rangle_i$ (doubly occupied). The physics of the
system is essentially that of $N_s=\sum_i c^\dagger_i c_i$
spinless fermions and $N_d$ bosons, with eigenstates given by \be
|\psi (N_s,N_d)> = \mathcal{N} (\eta^\dagger)^{N_d} a^\dagger_{0}
\cdots a^\dagger_{N_s-1}\ket{vac} \; . \label{psi}\ee Here
$\mathcal{N}=\Bigl[(L-N_s-N_d)!/(L-N_s)!N_d!\Bigr]^{1/2}$ is a
normalization factor; $a_j^\dagger$ is the Fourier transform of
the spinless fermion operator $c_j^\dagger$, $a_j^\dagger= \sum_q
{1\over \sqrt{L}}\exp(i { \pi\over L} j q) c^\dagger_{q}$, with
$j=0,\dots,N_s-1$; moreover $\eta^\dagger =\sum_{i=1}^L
\eta_i^\dagger$ is also known as the eta operator, and creates
doubly occupied sites from empty ones
($\eta_i^\dagger|0\rangle_i=|2\rangle_i$); $\ket{vac}$ is the
electron vacuum. $(\eta^\dagger)^{k}\ket{vac} $ is known to carry
ODLRO and multipartite entanglement \cite{vedraleta}. At fixed
filling $n=(N_s+2 N_d)/L$, the actual value of $N_s$ in
(\ref{psi}) is chosen to minimize the corresponding eigenvalue
$E(N_s,N_d)= -2  \frac {L}{\pi} \sin (\pi\frac{N_s}{L}) +u N_d$.

The system density matrix in the ground state is defined by
$\rho\doteq|\psi (N_s,N_d)><\psi (N_s,N_d)|$. Results of the
calculation for the single-site $\rho_i$ and the dimer $\rho_{ij}$
reduced density matrices are reported below. With respect to the
basis $|0\rangle$, $|\mbox{\o}\rangle$, $|2\rangle$, $\rho_i=
diag\{1-n_s-n_d,n_s,n_d\}$ with $n_\alpha\doteq {N_\alpha\over L}$
($\alpha=s,d$).  Whereas with respect to the basis $\ket{00}$,
$\ket{0\mbox{\o}}$, $\ket{\mbox{\o}0}$,
$\ket{\mbox{\o}\mbox{\o}}$, $\ket{\mbox{\o}2}$,
$\ket{2\mbox{\o}}$, $\ket{02}$, $\ket{20}$, $\ket{22}$,

\be \rho_{ij}= \left( \begin{array}{ccccccccc}
D_1  & 0      & 0    & 0    & 0      & 0    & 0  & 0  & 0    \\
0    & O_1    & O_2  & 0    & 0      & 0    & 0  & 0  & 0    \\
0    & O_2^*  & O_1  & 0    & 0      & 0    & 0  & 0  & 0    \\
0    & 0      & 0    & D_2  & 0      & 0    & 0  & 0  & 0    \\
0    & 0      & 0    & 0    & P_1    & P_2  & 0  & 0  & 0    \\
0    & 0      & 0    & 0    & P_2^*  & P_1  & 0  & 0  & 0    \\
0    & 0      & 0    & 0    & 0      & 0    & Q  & Q  & 0    \\
0    & 0      & 0    & 0    & 0      & 0    & Q  & Q  & 0    \\
0    & 0      & 0    & 0    & 0      & 0    & 0  & 0  & D_3  \\
\end{array} \right)
\ee Here, assuming $\ep=1/L$,

\[\begin{array}{cclccclcc}
D_1 & = & P_{ij} \frac{(\delta_s-n_d)(\delta_s-n_d-\ep)}{\delta_s
(\delta_s-\ep)}\quad & & \quad  O_2 &=& C_{ij}\frac{\delta_s-n_d}{\delta_s}\\
D_2 & = & P_{ij}+1-2 \delta_s   \quad & & \quad P_1 &=&
\frac{n_d}{\delta_s}\left( \delta_s -P_{ij}\right )\\
D_3 &=& \frac{n_d (n_d-\ep)}{\delta_s (\delta_s-\ep)} P_{ij}\quad
& & \quad P_2 &=& \frac{n_d}{\delta_s} C_{ij}\\
O_1 &=& \left(\delta_s-P_{ij}\right)
\left(\frac{\delta_s-n_d}{\delta_s}\right)\quad & & \quad Q &=&
\frac{n_d(\delta_s-n_d)}{\delta_s(\delta_s-\ep)}P_{ij}\\
 \end{array}  \label{entries}\]

\noindent with $P_{ij}=\delta_s^2-|C_{ij}|^2$,  $\delta_s=1-n_s$,
$|C_{ij}|=\ep \frac{\sin (n_s\pi |i-j|)}{\sin(\pi \ep |i-j|)}$. In
the thermodynamic limit $\epsilon \rightarrow 0$, $n_\alpha$
 finite, the above results may also be derived from \cite{SCHAD}.
\paragraph{Results.}
As a preliminary observation let us notice that in phases $I$,
$I'$ and $III$ (see Fig. \ref{3D}, left side) the dimension of
on-site vector space reduces to two, meaning that in these cases
$\mathcal{N}_{i,j}$ should reproduce results evaluated through
concurrence. This happens to be the case; in particular, in phase
$III$ $\mathcal{N}_{i,j}$ (and the concurrence) are vanishing
$\forall |i-j|$, whereas $\mathcal{I}_{i,j}$ is equal to $n
(2-n)/2$, which is related to the value of the ODLRO
parameter, in agreement with \cite{vedraleta}.\\
We also observe that whenever $C_{i,j}$ is zero (for instance,
phases $III$ and $IV$) the two-site density matrix is independent
of the sites $i$ and $j$. In the insulating phase $IV$ this
happens because the state is a tensor product of identical
single-site states, and all correlations are identically zero. On
the contrary, in phase $III$, since $Q\neq 0$, the surviving
two-point quantum correlations are range independent. In such
cases it may be useful to introduce global (\textit{i.e.} sums
over all sites) quantities instead of local ones, since it may
happen that a correlation is locally vanishing but globally
relevant; also finite size corrections ($\epsilon$) have to be
considered.
 For instance, it turns out that in so doing in phase
$III$ the total negativity becomes nonvanishing $\sum_{j\neq i}
\mathcal{N}_{i,j}= n (2-n)/[(2-n)^2 +n^2]$.

We now turn to discuss what happens at QPTs by studying the behavior of
$\mathcal{S}_i$, $\mathcal{I}_{i,j}$, $\mathcal{N}_{i,j}$. As mentioned, each of
the observed measures of correlations keeps track of the
undergoing transitions, exhibiting a singular behavior at the
transition points. The latter can be characterized by the analysis
of the partial derivatives of each measure. As an example, in the
right part of Fig. \ref{3D} we plot $\partial_u\mathcal{S}_{i}$.
Noticeably, the divergencies in the derivative are in perfect
correspondence with the parameter's values at which the various
QPTs occur, aside from transitions $I,I'\rightarrow IV$ that must
be revealed by $\partial_n\mathcal{S}_{i}$. The systematic
analysis of the behavior of the various derivatives at each QPT is
carried on in Table \ref{table}.
 \begin{table}[h]
\begin{center}
\begin{tabular}{|l||c|c|c|c|}
\hline
 & $\partial_x\mathcal{S}_{i}$& $\partial_x\mathcal{I}_{i,j}$ & $\partial_x\mathcal{N}_{i,j}$  & ent\\
\hline \hline
 $I,I' \rightarrow IV (x=n)$
        & $\log|n_c-n|$&  FD &  $0$  & Multi  \\

\hline
 $II\rightarrow I,I' (x=u)$
        & $\log(u_c-u)$&  FD &  FD & Multi   \\

\hline $II\rightarrow I,I' (x=n)$
        & $\log|n-n_c|$ &  FD &  FD & Multi  \\
\hline $II\rightarrow III (x=u)$
        &  $1/\sqrt{u-u_c}$  & $1/\sqrt{u-u_c}$  & FD & Two\\

\hline $II \rightarrow IV (x=u)$
        &  $1/\sqrt{u_c-u}$ & $1/\sqrt{u_c-u}$  & FD & Two\\

\hline
\end{tabular}
\end{center}
\caption{\small{Behavior of the evaluated partial derivatives at
critical points for the various QPTs (left column): FD is finite
discontinuity, ``multi'' refers to multipartite, and ``two''
refers to two-point.}} \label{table}
\end{table}
We first consider $\partial_x\mathcal{S}_{i}$, $x=n,u$; it
exhibits two different kinds of divergencies: logarithmic for
transitions $I,I'\rightarrow IV$ and $II\rightarrow I,I'$;
algebraic for transitions $II\rightarrow III$ and $II\rightarrow
IV$, with exponent $\nu=1/2$. The latter turns out to correspond
to the shift exponent as extracted from finite size analysis
\cite{AGM}.
\\In Tab. \ref{table} we also report the
behavior of $\partial_x\mathcal{I}_{i,j}$ and
$\partial_x\mathcal{N}_{i,j}$ at QPTs. As described in the
paragraph \textit{Entanglement and QPTs}, the comparison of the
three quantities can be used to understand whether bipartite or
multipartite entanglement is relevant to the various transitions.
In fact, all transitions corresponding to a (logarithmic)
divergence in $\partial_x \mathcal{S}_i$ are not seen as
divergencies either in $\partial_x \mathcal{I}_{i,j}$ or in
$\partial_x \mathcal{N}_{i,j}$. In such cases, we infer that the
transitions are to ascribe to $QS$ correlations; this is also in
agreement with the fact in some of these transitions ($II
\rightarrow I,I'$) the component of the ground state given by the
eta pairs (which carry multipartite entanglement and ODLRO)
disappears.  On the contrary, whenever the divergent behavior
exhibited by $\partial_x \mathcal{S}_i$ is also displayed by
$\partial_x \mathcal{I}_{i,j}$, as seen for the two transitions
$II\rightarrow III$ and $II\rightarrow IV$, this is to be
interpreted as a signal of the role of $Q2$ correlations in the
QPT.

Different from what happens for the first two quantities reported
in Table \ref{table}, we can check on the third column that
$\partial_x \mathcal{N}_{i,j}$ never does display the same
singular behavior as $\partial_x S_{i}$. In particular, this
happens in correspondence of the transitions driven by two-point
correlations, $II\rightarrow III$ and $II\rightarrow IV$,
suggesting that $\mathcal{N}_{i,j}$ is just a lower bound for $Q2$
correlations also for the present model. Apart from this fact, the
behavior of $\mathcal{N}_{i,j}$ for various values of $|i-j|$
supports once more the idea that the transitions in question have
to be ascribed to two-point correlations.
\begin{figure}[h]
\psfrag{N}{$\mathcal{N}_{i,j}$} \psfrag{I}{$\mathcal{I}_{i,j}$}
 \psfrag{S}{$\mathcal{S}_i$}
\psfrag{D}{$\mathcal{S}_{i,j}$}
 \psfrag{u}{\small $u$}
 \psfrag{a}{1}
 \psfrag{b}{2}
 \psfrag{c}{3}
 \psfrag{d}{4}
 \psfrag{e}{5}
\hbox{\includegraphics[height=6cm, width=8.57cm, viewport= 0 20
560 680, clip]{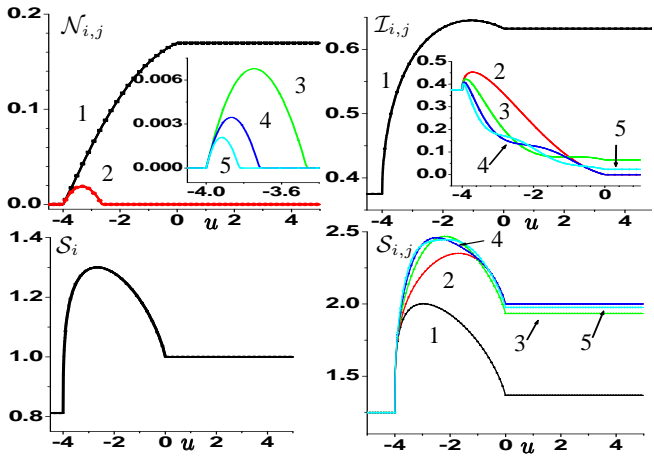}} \caption{\small{Plots of
$\mathcal{N}_{i,j}$, $\mathcal{I}_{i,j}$, $\mathcal{S}_i$ and
$\mathcal{S}_{i,j}$ ($|i-j|=1,...,5$) for the section $n=0.5$
(line $\alpha$ in Fig. \ref{3D})}}
 \label{dfix}
\end{figure}
As an example, we report in the top left part of Fig. \ref{dfix}
$\mathcal{N}_{i,j}$ for the case $n=0.5$; the transition
$II\rightarrow III$ takes place at $u_c=-4$. As $u$ gets close to
$u_c$ two-point quantum correlations begin to spread along the
chain; this is shown by the non zero value of $\mathcal{N}_{i,j}$
for an increasing number of pairs of sites whose distance $|i-j|$
grows up to $\infty$ as $u \rightarrow u_c$. This is a clear
indication of diverging correlation length originated from $Q2$
correlations at critical point. One could expect that again the
total negativity is the right quantity to display a critical
behavior, in agreement with similar conclusions about concurrence
\cite{SWHL} in spin-$1/2$ systems. Moreover, the value at which
$\mathcal{N}_{i,j}$ reaches its maximum gets closer to $u_c$ by
increasing $|i-j|$, indicating its possible scaling behavior. The
same qualitative behavior of maximum is observed for
$\mathcal{I}_{i,j}$ (top right part of figure), even though such a
quantity is, in general, different from zero also away from the
critical point, suggesting once more that the quantum mutual
information in the vicinity of the transition captures the
divergent behavior of just the $Q2$ correlations. We finally
analyze in Fig. \ref{dfix}
$\mathcal{S}_{i,j}=\mathcal{S}(\rho_{i,j})$, which describes all
quantum correlations between the dimer $i,j$ and the rest of the
system. Interestingly, $\mathcal{S}_{i,j}$ has for all $i,j$ the
same qualitative behavior of $\mathcal{S}_i$ at critical points.
Such a feature is confirmed by our calculations in correspondence
of all QPTs. This is expected within our scheme, since $S_i$ and
$\mathcal{S}_{i,j}$ both describe the same correlations, Q2 and
QS.

\paragraph{Conclusions.} We have studied
the behavior of different measures of correlations in
correspondence of QPTs for a solvable model of correlated
electrons on a chain€, displaying different kinds of
metal-insulator-superconductor transitions. As a general output of
our work, the role of quantum mutual information in the
investigation of QPTs has been recognized. In particular, the
comparison of singularities of the latter quantity with
singularities of single-site entanglement allows one to
distinguish at each QPT the contribution of bipartite from that of
multipartite entanglement. At the same time, whenever a
contribution from two-point quantum correlations is spotted, this
can be used to test direct measures of bipartite entanglement. As
an example, we tested the negativity, finding that in this case it
does not capture all of the two-point quantum correlations, though
it shows evidence of a diverging correlation length and
interesting scaling behavior in the vicinity of the transitions
ascribed to two-point correlations. The study of scaling
properties of the proposed measures of entanglement and of total
negativity \cite{AGM} need to be further investigated.

\end{document}